\documentclass[aps,pra,twocolumn,showpacs,superscriptaddress]{revtex4-1}
\usepackage{amsmath,amssymb}
\usepackage{graphicx}
\usepackage{natbib}
\usepackage{bm}
\usepackage{braket}

\begin{document}
\title{Magnetic resonance spectroscopy and characterization of magnetic
phases for spinor Bose-Einstein condensates}
\author{Akiyuki Tokuno}
\affiliation{DPMC-MaNEP, University of Geneva, 24 Quai Ernest-Ansermet,
CH-1211 Geneva, Switzerland.}
\author{Shun Uchino}
\affiliation{DPMC-MaNEP, University of Geneva, 24 Quai Ernest-Ansermet,
CH-1211 Geneva, Switzerland.}

\date{\today}

\begin{abstract}
 The response of spinor Bose-Einstein condensates to dynamical
 modulation of magnetic fields is discussed with linear response
 theory. 
 As an experimentally measurable quantity, the energy absorption rate
 (EAR) is considered, and the response function is found to access
 quadratic spin correlations which come from the perturbation of the
 quadratic Zeeman term.
 By applying our formalism to spin-$1$ condensates, we demonstrate that
 the EAR spectrum as a function of the modulation frequency is able to
 characterize the different magnetically ordered phases.
\end{abstract}

\maketitle

\textit{Introduction.---}
Ultracold Bose atoms with spin degrees of
freedom~\cite{Ho/PRL81.1998,Ohmi.Machida/JPSJ67.1998,Stamper-Kurn.etal/PRL80.1998,Barrett.etal/PRL87.2001,Ueda/AnnRevCondMattPhys3.2012:Review,Kawaguchi.Ueda/PhysRep520.2012:Review,Stamper-Kurn.Ueda/arXiv2012:Review}
have been attracting interest as a class of quantum fluids accompanying
nontrivial spin orders and topological spin textures, in contrast with 
spinless Bose-Einstein condensates (BECs).
In BECs with spins, so-called spinor BECs, spin rotational symmetry
allows spin-dependent interactions, and the number of the independent 
interactions increases with spin degrees of freedom of atoms, which
causes various ground states.
However, in addition to exploring the properties of those nontrivial
spin orders, it is also important to specify their equilibrium
properties experimentally. 
Thus, the development of measurement techniques to capture the physical
properties of complicated ordered states is a challenge for the study of
spinor BECs. 

The powerful way to identify the mean-field ground state and the
phase diagrams is the measurement of population of the spin components by
the combination of the Stern-Gerlach experiment and time-of-flight (TOF)
analysis.~\cite{Stenger.etal/Nature396.1998,Black.etal/PRL99.2007,Kuwamoto.etal/PRA69.2004,Chang.etal/PRL92.2004,Schmaljohann.etal/PR92.2004,Pasquiou.etal/PRL106.2011}
In addition, the dispersive imaging method with off-resonant light allows
for displaying spatially resolved spin
profiles~\cite{Carusotto.Mueller/JPhysB37.2004,Higbie.etal/PRL95.2005,Liu.etal/PRL102.2009,Kuzmich.etal/PRA60.1999}. 
At the same time, while the current techniques probe equilibrium properties,
it is also challenging to provide more
direct and systematic probes to visualize the excitation energy structure
coming from spin fluctuations.

For spinor BECs in the presence of uniform magnetic fields, the
quadratic Zeeman (QZ) shift, in addition to the linear Zeeman (LZ) shift,
emerges due to hyperfine couplings between a nuclear and an electron
spin~\cite{Stenger.etal/Nature396.1998}.
Because the magnetization of spinor BECs is known to be preserved at
least within the limit of accuracy of experimental
errors~\cite{Chang.etal/PRL92.2004}, the QZ shift is the most relevant
effect induced by the magnetic field.
In addition and importantly, the QZ coupling is experimentally
controllable~\cite{Gerbier.etal/PRA73.2006,Leslie.etal/PRA79.2009,Bookjans.Vinit.Raman/PRL107.2011,Jacob.etal/PRA86.2012,Sadler.etal/Nature443.2006,Guzman.etal/PRA84.2011}.

In this Rapid Communication, motivated by such a possible control of the QZ coupling,
we consider magnetic resonant spectra as a response to dynamically
modulated magnetic fields, which possesses the potential to probe
microscopic spin-excitation energy structures.
As a measurable quantity, we focus on the energy absorption rate (EAR), and
formulate it with linear response theory.
The consequent formula is applicable to general systems with any spin
degree of freedom.
This type of resonance spectra has not been considered, and thus
it is important to clarify the spectral features for various states.
As a simple case, we take spin-$1$ BECs in this Rapid Communication and
calculate the spectrum with Bogoliubov
theory~\cite{Murata.Saito.Ueda/PRA75.2007,Uchino.Kobayashi.Ueda/PRA81.2010}.
As a result, they are found to exhibit different behaviors in each phase. 
Furthermore, we also consider the cases in the presence of trap
potentials and a noncondensed fraction, and the ordered states are
concluded to remain distinguishable from the low-frequency behaviors of
the EAR spectra.

\textit{Formalism.---}
We start with general spin-$F$ Bose atom systems under a uniform
magnetic field. 
Let us suppose the many-body static Hamiltonian including Zeeman
couplings to be $H_{0}$.
In this Rapid Communication, we restrict ourselves to the cases for
which the magnetic field is applied along the $z$ axis, and $H_0$ is
invariant under spin rotation around the $z$ axis, which is a general
setup in experiments. 

In the presence of a dynamically modulated magnetic field such as
$h+\delta{h}\cos(\omega t/2)$, the system should be described by the
time-dependent Hamiltonian $H(t)=H_{0}+V(t)$, and the perturbation 
is represented as
\begin{equation}
 V(t)
 =(\delta{p})\cos(\omega t/2)\mathcal{F}_{\mathrm{L}}
  +(\delta{q})\cos^2(\omega t/2)\mathcal{F}_{\mathrm{Q}},
 \label{eq:perturbation}
\end{equation}
where the first and second terms mean modulation of the LZ and QZ
couplings, respectively.
The coupling constants, $\delta{p}$ and $\delta{q}$, are proportional
to $\delta{h}$ and $(\delta{h})^2$, respectively.
The LZ and QZ operators are represented as
\begin{align}
 &\mathcal{F}_{\mathrm{L}}
 =\int\!\!d\bm{r}\,
  \hat{\Psi}^{\dagger}(\bm{r})F^{z}\hat{\Psi}(\bm{r}),
 \\
 &\mathcal{F}_{\mathrm{Q}}
 =\int\!\!d\bm{r}\, 
  \hat{\Psi}^{\dagger}(\bm{r})(F^{z})^2\hat{\Psi}(\bm{r}),
\end{align}
where
$\hat{\Psi}=(\psi_{F},\psi_{F-1},\cdots,\psi_{-F})^{T}$
denotes a spinor boson field, and $\bm{F}=(F^{x},F^{y},F^{z})$ is a
spin-$F$ matrix.

In experiments, the EAR can be measured through the TOF image.
Assuming the energy scale of the periodically modulated perturbation 
is small enough~\footnote{For example, the modulation frequency
$\hbar\omega$ would be sufficiently smaller than mean-field condensation
energy.}, the dynamics is well described with linear response theory.
Then, the EAR is defined as 
$R(\omega)=\frac{1}{2\pi/\omega}\int_{T}^{T+2\pi/\omega}\!\!d{t}\frac{d\braket{H(t)}}{dt}$,
where $\braket{\cdots}$ denotes the statistical average over $H(t)$. 
Thus, the EAR is derived as
\begin{equation}
 R(\omega)
 =-\frac{1}{2\hbar}
  \omega\mathrm{Im}[\chi^\mathrm{R}(\omega)],
 \label{eq:EAR}
\end{equation}
where 
$\chi^{\mathrm{R}}(\omega)=-i\int_{0}^{\infty}dt\,e^{i\omega t}\braket{[V(t),V(0)]}_0$ 
is the retarded correlation function of the
perturbation~(\ref{eq:perturbation}) averaged over $H_0$.
Since $H_0$ is assumed to possess the spin rotational symmetry around
the $z$ axis, 
$[\mathcal{F}_{\mathrm{L}},H_0]=[\mathcal{F}_{\mathrm{L}},\mathcal{F}_{\mathrm{Q}}]=0$,
and thus the retarded correlation function is reduced to
\begin{equation}
 \chi^{\mathrm{R}}(\omega)
 =-i(\delta{q})^2\int_{0}^{\infty}\!\!d{t}\,
  e^{i\omega t}
  \braket{[\mathcal{F}_{\mathrm{Q}}(t),\mathcal{F}_{\mathrm{Q}}(0)]}_0.
\end{equation}
Namely, the system is insensitive to the dynamic modulation of the LZ
coupling.
The remarkable point here is that the obtained formula is generic, and
applicable for any spin degrees of freedom and form of interactions, as
long as the uniaxial spin rotational symmetry exists at least.

\textit{EAR for spin-$1$ BECs.---}
Let us demonstrate the EAR spectrum~(\ref{eq:EAR}) to allow for
characterizing spin-ordered phases.
We consider spin-$1$ interacting
bosons~\cite{Ho/PRL81.1998,Ohmi.Machida/JPSJ67.1998,Stenger.etal/Nature396.1998}
without a trap, which undergo a BEC in the low-temperature regime .
Hereafter, we fix total spin to be zero.~\footnote{The ferromagnetic
state breaks this assumption, but it then means the formation of the
ferromagnetic domain structure, and the calculated EAR spectrum
corresponds to that from the bulk domains.} 
Then, since the LZ term is effectively
vanished~\cite{Stenger.etal/Nature396.1998}, the Hamiltonian to be
considered is given by
\begin{align}
 H_0
 &
 =\int\!\!d\bm{r}\,
  \biggl[
   -\frac{\hbar^2}{2M}\hat{\Psi}^{\dagger}(\bm{r})\nabla^2\hat{\Psi}(\bm{r})
   +q\hat{\Psi}^{\dagger}(\bm{r})(F^z)^2\hat{\Psi}(\bm{r})
 \nonumber \\
 &\quad
   +\frac{c_0}{2}
    \Bigl(\hat{\Psi}^{\dagger}(\bm{r})\hat{\Psi}(\bm{r})\Bigr)^2
   +\frac{c_1}{2}
    \Bigl(\hat{\Psi}^\dagger(\bm{r})\bm{F}\hat{\Psi}(\bm{r})\Bigr)^2
  \biggr],
 \label{eq:spinor-Hamiltonan}
\end{align}
where $\hat{\Psi}=(\psi_{1},\psi_{0},\psi_{-1})^{T}$, 
$M$ denotes the mass of the atoms, and $c_0$ and $c_1$ mean the density
and spin exchange interactions, respectively.
For $^{23}$Na and $^{87}$Rb atoms, the coupling $c_1$ is taken to be
a positive and negative value, respectively. 
Let us impose $n c_0\gg n |c_1|, |q|$, where $n$ is the atom density
corresponding to the experimental conditions. 
Since the EAR is independent of the LZ coupling modulation, as discussed
above, instead of (\ref{eq:perturbation}) we here suppose the modulation
perturbation of the magnetic field to be
\begin{equation}
 V(t)=(\delta{q})\cos(\omega t)\mathcal{F}_{\mathrm{Q}}.
\end{equation}

\begin{figure}[tbp]
 \includegraphics[scale=.7]{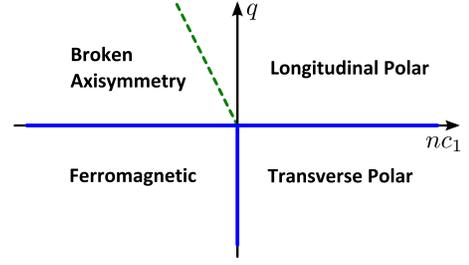}
 \caption{(Color online)
 The MF phase diagram of spin-$1$ BECs with respect to the effective
 spin-exchange interaction $n c_1$ and QZ coupling $q$, which are here
 restricted to being weak compared with the density interaction, i.e.,
 $|q|,\,n|c_1|\ll nc_0$.
 The thick-solid and dashed boundary lines denote first- and second-order
 phase transitions, respectively.
 We note that the phase diagram was first given in
 \cite{Stenger.etal/Nature396.1998}, which covers a wider parameter
 regime than shown here.
 The phase boundary between the BA and LP phase is given by
 $q=2n|c_1|$. 
 }
 \label{fig:phase-diagram}
\end{figure}

The mean-field (MF) analysis, in which the field $\hat{\Psi}$ is
replaced by a MF spinor order parameter $\hat{\xi}$ optimizing the
Hamiltonian~(\ref{eq:spinor-Hamiltonan}),
leads to the following ground
states~\cite{Uchino.Kobayashi.Ueda/PRA81.2010}~\footnote{For simplicity,
the relative phases of the MF spinors are fixed to be zero here, which does
not affect the result shown in this Rapid Communication.}
as shown in Fig.~\ref{fig:phase-diagram}:
(i) the ferromagnetic (FM) phase
$\hat{\xi}_{\mathrm{FM}}=(1,0,0)^{T}$ for $c_1,q<0$, 
(ii) the longitudinal polar (LP) phase 
$\hat{\xi}_{\mathrm{LP}}=(0,1,0)^{T}$ for $q>0$ and $q+2nc_1>0$,
(iii) the transverse polar (TP) phase 
$\hat{\xi}_{\mathrm{TP}}=(1/\sqrt{2},0,1/\sqrt{2})^{T}$ for $c_1>0$ and
$q<0$,
and (iv) the broken-axisymmetry (BA) phase
$\hat{\xi}_{\mathrm{BA}}=(\sin{\theta}/\sqrt{2},\cos{\theta},\sin{\theta}/\sqrt{2})^{T}$,
where $\sin{\theta}=\sqrt{(1-\tilde{q})/2}$ with
$\tilde{q}=q/(2n|c_1|)$, for $c_1<0$ and $0<q<2n|c_1|$.

The correlation function $\chi^{\mathrm{R}}(\omega)$ is
calculated with the Bogoliubov theory by the replacement
$\hat{\Psi}=\sqrt{N_0}\hat{\bm{\xi}}_{\alpha}+\delta{\hat{\Psi}}$ 
($\alpha=\mathrm{FM},\mathrm{LP},\mathrm{TP},\mathrm{BA}$), where $N_0$
is the condensate atom number. 
Then the Bogoliubov Hamiltonian is represented as
$H_0\approx E_{\mathrm{MF}}+H_{\mathrm{eff}}$, with the MF energy
$E_{\mathrm{MF}}$ and
\begin{align}
 H_{\mathrm{eff}}
 &=\sum_{\bm{k}\ne\bm{0}}
   \biggl[
   \hat{a}^{\dagger}_{\bm{k}}
   \Bigl(
     \epsilon_{\bm{k}}
     +nc_1\overline{\bm{F}}\cdot(\bm{F}-\overline{\bm{F}})
     +q((F^z)^2-\overline{(F^z)^2})
   \Bigr)
   \hat{a}_{\bm{k}}
 \nonumber \\
 &\quad
  +\frac{nc_0}{2}
   \Bigl(
    D^{\dagger}_{\bm{k}}D_{\bm{k}}
    +D_{\bm{k}}D_{-\bm{k}}
    +\mathrm{H.c}
   \Bigr)
 \nonumber \\
 &\quad
 +\frac{nc_1}{2}
  \Bigl(
   \bm{S}^{\dagger}_{\bm{k}}\cdot\bm{S}_{\bm{k}}
   +\bm{S}_{\bm{k}}\cdot\bm{S}_{-\bm{k}}
   +\mathrm{H.c}
  \Bigr)
  \biggr],
\end{align}
where $\epsilon_{\bm{k}}=\frac{\hbar^2\bm{k}^2}{2M}$, and 
$\hat{a}_{\bm{k}}=\int\!\!d\bm{r}e^{i\bm{k}\cdot\bm{r}}\delta{\hat{\Psi}}=(a_{1\bm{k}},a_{0\bm{k}},a_{-1\bm{k}})^{T}$
is a Fourier transform of the spinor fluctuation.
$D_{\bm{k}}=\hat{\xi}^{\dagger}\hat{a}_{\bm{k}}$ and
$\bm{S}_{\bm{k}}=\hat{\xi}^{\dagger}\bm{F}\hat{a}_{\bm{k}}$ 
mean density and spin fluctuations, respectively, and
$\overline{X}=\hat{\xi}^{\dagger}X\hat{\xi}$, for a spin operator $X$,
denotes the MF average of spin matrices.
In this representation, the QZ operator is expressed as 
\begin{equation}
 \mathcal{F}_{\mathrm{Q}}
 =N\overline{(F^{z})^{2}}
  +\sum_{\bm{k}\ne\bm{0}}
   \hat{a}^{\dagger}_{\bm{k}}
   \left[(F^{z})^2-\overline{(F^{z})^{2}}\right]
   \hat{a}_{\bm{k}},
\end{equation}
where $N$ denotes the atom number.

\textit{FM phase.---}
The MF spinor leads to the density and spin fluctuation as
$D_{\bm{k}}=S^{z}_{\bm{k}}=a_{1\bm{k}}$,
$S^{x}_{\bm{k}}=a_{0\bm{k}}/\sqrt{2}$, and
$S^{y}_{\bm{k}}=ia_{0\bm{k}}/\sqrt{2}$.
The effective Hamiltonian diagonalized by the Bogoliubov transformation 
is given~\cite{Uchino.Kobayashi.Ueda/PRA81.2010} as 
\begin{align}
 H^{\mathrm{FM}}_{\mathrm{eff}}
 &=\sum_{\bm{k}\ne\bm{0}}
   \biggl[
    E^{\mathrm{FM}}_{\mathrm{d}}(\bm{k})
    d^{\dagger}(\bm{k})d(\bm{k})
    +E^{\mathrm{FM}}_{z}(\bm{k})f^{\dagger}_{z}(\bm{k})f_{z}(\bm{k})
 \nonumber \\
 &\quad
    +E^{\mathrm{FM}}_{xy}(\bm{k})f^{\dagger}_{xy}(\bm{k})f_{xy}(\bm{k})
   \biggr],
 \label{eq:Bogoliubov-Hamiltonian-FM}
\end{align}
where
$E^{\mathrm{FM}}_{\mathrm{d}}(\bm{k})=\sqrt{\epsilon_{\bm{k}}[\epsilon_{\bm{k}}+2n(c_0-|c_1|)]}$,
$E^{\mathrm{FM}}_{z}(\bm{k})=\epsilon_{\bm{k}}+2n|c_1|$, and
$E^{\mathrm{FM}}_{xy}(\bm{k})=\epsilon_{\bm{k}}+|q|$.
The QZ operator in the FM phase is written as
\begin{align}
 \mathcal{F}^{\mathrm{FM}}_{\mathrm{Q}}
 =\mathrm{Const.}
  +\sum_{\bm{k}\ne\bm{0}}f^{\dagger}_{xy}(\bm{k})f_{xy}(\bm{k}).
 \label{eq:QZ-FM}
\end{align}
Since the perturbation $\mathcal{F}^{\mathrm{FM}}_{\mathrm{Q}}$
commutes with the Hamiltonian~(\ref{eq:Bogoliubov-Hamiltonian-FM}),
$\chi^{\mathrm{R}}(\omega)$ is immediately found to be zero. 
Thus, the EAR spectrum shows no signal in the entire $\omega$ regime.

\begin{figure*}[tbp]
 \begin{center}
  \includegraphics[scale=.37]{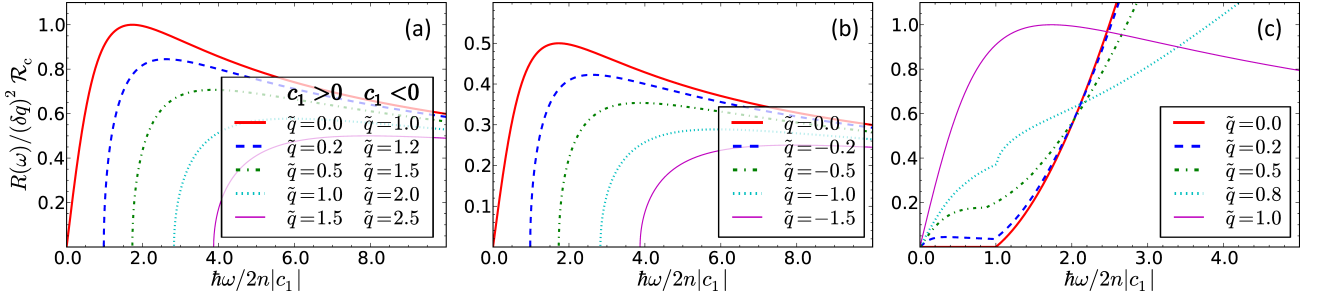}
  \caption{(Color online)
  The EAR spectra as a function of modulation frequency for the
  different $\tilde{q}=q/2n|c_1|$: 
  (a) the LP phase for $c_1, q>0$ or for $q>0$ and $q+2nc_1>0$,
  (b) the TP phase for $c_1>0$ and $q<0$, 
  and (c) the BA phase for $c_1<0$ and $0<q<2n|c_1|$.
  Spectra (a) and (b) are qualitatively the same, but
  the intensity differs due to the different number
  of accessible gapful-spin modes. 
  Spectra (c) apart from $\tilde{q}=1$ show an abrupt enhancement
  of the EAR around $\hbar\omega\approx 2n|c_1|$, which indicates the
  appearance of the contribution from
  $r_{\mathrm{BA}_\mathrm{d}}(\omega)$.
  }
  \label{fig:EAR}
 \end{center}
\end{figure*}

\textit{LP phase.---}
The diagonalized Bogoliubov Hamiltonian is
given (see~\cite{Uchino.Kobayashi.Ueda/PRA81.2010} and Supplemental
Material~\cite{sm}) as 
\begin{align}
 H^{\mathrm{LP}}_{\mathrm{eff}}
 =\sum_{\bm{k}\ne\bm{0}}
  \left[
    E_{\bm{k}}d^{\dagger}(\bm{k})d(\bm{k})
    +\sum_{\nu=x,y}
       E^{f}_{\bm{k}}f^{\dagger}_{\nu}(\bm{k})f_{\nu}(\bm{k})
  \right],
 \label{eq:Hamiltonian-P}
\end{align}
where $E_{\bm{k}}=\sqrt{\epsilon_{\bm{k}}(\epsilon_{\bm{k}}+2nc_0)}$
and
$E^{f}_{\bm{k}}=\sqrt{(\epsilon_{\bm{k}}+|q|)(\epsilon_{\bm{k}}+|q|+2nc_1)}$
are, respectively, a gapless-phonon and doubly degenerate gapful-spin
mode. 
The Bogoliubov transformation gives the form of the QZ
perturbation (see Supplemental Material~\cite{sm}) as 
\begin{align}
 \mathcal{F}^{\mathrm{LP}}_{\mathrm{Q}}
 &=\bar{\mathcal{F}}^{\mathrm{LP}}_{\mathrm{Q}}
   -\sum_{\bm{k}\ne\bm{0}}\sum_{\nu=x,y}
    \mathcal{A}^{\mathrm{LP}}_{\bm{k}}
     \biggl[
       f^{\dagger}_{\nu}(\bm{k})f^{\dagger}_{\nu}(-\bm{k})
       +\mathrm{H.c.}
     \biggr],
\label{eq:QZ-polar1}
\end{align}
where $\bar{\mathcal{F}}^{\mathrm{LP}}_{\mathrm{Q}}$ contains terms
which commute with the Hamiltonian~(\ref{eq:Hamiltonian-P}), and
$\mathcal{A}^{\mathrm{LP}}_{\bm{k}}=n|c_1|/2E^{f}_{\bm{k}}$.
The QZ perturbation accesses the two spin modes.
From Eq.~(\ref{eq:QZ-polar1}), the retarded correlation function is
straightforwardly calculated, and consequently the EAR spectrum is
analytically obtained (see Supplemental Material~\cite{sm}) as 
$R_{\mathrm{LP}}(\omega)=(\delta{q})^2\mathcal{R}_{\mathrm{c}}\theta_{\mathrm{H}}(|\tilde{\omega}|-2\tilde{\Delta}_{q})r_{\mathrm{LP}}(\omega)$ 
with 
\begin{align}
 r_{\mathrm{LP}}(\omega)
 =2
   \sqrt{\frac{\sqrt{1+\tilde{\omega}^2}-2\tilde{q}-\mathrm{sgn}(c_1)}{1+\tilde{\omega}^2}},
 \label{eq:EAR-polar1}
\end{align}
where $\theta_{\mathrm{H}}(x)$ is a Heaviside step function, and
$\mathcal{R}_{\mathrm{c}}=\Omega(2Mn|c_1|)^{3/2}/64\pi\hbar^4$ with the
system volume $\Omega$ is a constant. 
We have taken
$\tilde{\omega}=\hbar\omega/2n|c_1|$, $\tilde{q}=q/2n|c_1|$, and 
$\tilde{\Delta}_{q}=\sqrt{|\tilde{q}|(|\tilde{q}|+\mathrm{sgn}(c_1))}$.
The spin gap in the polar phase has been denoted by
$2n|c_1|\tilde{\Delta}_{q}$.
Equation~(\ref{eq:EAR-polar1}) for various values of $q$ is plotted in
Fig.~\ref{fig:EAR}(a).
Note that the gap of the EAR spectrum closes on the phase boundaries,
$q=0$ with $c_1>0$ and $q=2n|c_1|$ with $c_1<0$.

\textit{TP phase.---}
The Bogoliubov transformation diagonalizes the
Hamiltonian (see~\cite{Uchino.Kobayashi.Ueda/PRA81.2010} and
Supplemental Material~\cite{sm} as 
\begin{align}
 H^{\mathrm{TP}}_{\mathrm{eff}}
 =\sum_{\bm{k}\ne\bm{0}}
    \left[
      E_{\bm{k}}d^{\dagger}(\bm{k})d(\bm{k})
      +E^{z}_{\bm{k}}f_{z}^{\dagger}(\bm{k})f_{z}(\bm{k})
      +E^{f}_{\bm{k}}f^{\dagger}(\bm{k})f(\bm{k})
    \right],
 \label{eq:Hamiltonian-polar2}
\end{align}
where $E_{\bm{k}}$ and
$E^{z}_{\bm{k}}=\sqrt{\epsilon_{\bm{k}}(\epsilon_{\bm{k}}+2nc_1)}$ are
gapless modes of phonon and spin, respectively, and $E^{f}_{\bm{k}}$ is
another spin mode with the gap $\tilde{\Delta}_{q}$. 
The QZ operator is written (see Supplemental Material~\cite{sm}) as
\begin{align}
 \mathcal{F}^{\mathrm{TP}}_{\mathrm{Q}}
 &=\bar{\mathcal{F}}^{\mathrm{TP}}_{\mathrm{Q}}
  -\sum_{\bm{k}\ne\bm{0}}
    \mathcal{A}^{\mathrm{TP}}_{\bm{k}}
    \biggl[
     f^{\dagger}(\bm{k})f^{\dagger}(-\bm{k})
     +\mathrm{H.c.}
    \biggr],
 \label{eq:QZ-polar2}
\end{align}
where $\bar{\mathcal{F}}^{\mathrm{TP}}_{\mathrm{Q}}$ denotes terms which
commute with the Hamiltonian~(\ref{eq:Hamiltonian-polar2}), and
$\mathcal{A}^{\mathrm{TP}}_{\bm{k}}=nc_{1}/2E^{f}_{\bm{k}}$. 
From Eq.~(\ref{eq:QZ-polar2}), the QZ modulation is found to access only
one gapful-spin mode.
Straightforwardly, the retarded correlation function of
Eq.~(\ref{eq:QZ-polar2}) is calculated and the EAR is analytically
obtained (see Supplemental Material~\cite{sm}) as
$R_{\mathrm{TP}}(\omega)=(\delta{q})^2\mathcal{R}_{\mathrm{c}}\theta_{\mathrm{H}}(|\tilde{\omega}|-2\tilde{\Delta}_{q})r_{\mathrm{TP}}(\omega)$ with
\begin{align}
 r_{\mathrm{TP}}(\omega)
 &=\sqrt{\frac{\sqrt{1+\tilde{\omega}^2}-2|\tilde{q}|-1}{1+\tilde{\omega}^2}},
 \label{eq:EAR-polar2}
\end{align}
which is illustrated in Fig.~\ref{fig:EAR}.

It is remarkable that for positive $c_1$, we have
$R_{\mathrm{LP}}(\omega)/R_{\mathrm{TP}}(\omega)=2$ for the fixed $|q|$,
and the factor $2$ is a robust number because it comes from the
accessible number of the spin modes by the QZ perturbation; 
namely, the perturbation~(\ref{eq:QZ-polar1}) for the LP phase
involves the two gapful-spin modes, while only one gapful-spin mode is
accessed for the TP phase.
Although the two polar phases just have a quantitative difference and
other calibration may be needed to explicitly differentiate them through
a single measurement under a certain $|q|$, the different spectral
intensity still has an interesting aspect: 
If we continuously change $q$ across $q=0$, a discontinuous spectrum
change is observed, and it would be interpreted to be a signal of the
first-order phase transition associated with the spontaneous symmetry
breaking between the two different polar directions.

In summary, the EAR in the polar phases has two important features: 
The first is that we can measure the spin gap $\tilde{\Delta}_{q}$ which
dominates the low-energy spin excitation.
The second is that the discontinuous difference of the spectral
intensity allows for observing the first order phase transition from the
dynamical viewpoint.
As we will discuss later, these conclusions do not change even if we
have trap potentials.

\textit{BA phase.---}
For $c_0\gg |c_1|$, the Bogoliubov Hamiltonian is
diagonalized
(see~\cite{Uchino.Kobayashi.Ueda/PRA81.2010,Murata.Saito.Ueda/PRA75.2007,Barnett.etal/PRA84.2011}
and Supplemental Material~\cite{sm}
as 
\begin{align}
 H^{\mathrm{BA}}_{\mathrm{eff}}
 &=\sum_{\bm{k}\ne\bm{0}}
   \biggl[
    E^{\mathrm{BA}_{\mathrm{d}}}_{\bm{k}}d^{\dagger}(\bm{k})d(\bm{k})
    +E^{\mathrm{BA}_z}_{\bm{k}}f^{\dagger}_{z}(\bm{k})f_{z}(\bm{k})
 \nonumber \\
 &\quad
    +E^{\mathrm{BA}_{xy}}_{\bm{k}}f^{\dagger}_{xy}(\bm{k})f_{xy}(\bm{k})
   \biggr],
 \label{eq:Hamiltonian-BA}
\end{align}
where
$E^{\mathrm{BA}_{\mathrm{d}}}_{\bm{k}}=\sqrt{\epsilon_{\bm{k}}[\epsilon_{\bm{k}}+2nc_0-2n|c_1|(1-\tilde{q}^2)]}$,
$E^{\mathrm{BA}_z}_{\bm{k}}=\sqrt{\epsilon_{\bm{k}}(\epsilon_{\bm{k}}+q)}$,
and
$E^{\mathrm{BA}_{xy}}_{\bm{k}}=\sqrt{(\epsilon_{\bm{k}}+2n|c_1|)[\epsilon_{\bm{k}}+2(1-\tilde{q}^2)n|c_1|]}$
are interpreted to be the density mode, the gapless-spin mode and the
gapful-spin mode, respectively. 
The QZ perturbation is represented (see Supplemental Material~\cite{sm})
as
\begin{align}
 \mathcal{F}_{\mathrm{Q}}^{\mathrm{BA}}
 &=\bar{\mathcal{F}}^{\mathrm{BA}}_{\mathrm{Q}}
   +\sum_{\bm{k}\ne\bm{0}}
     \biggl[
      -\mathcal{A}^{\mathrm{BA}}_{\bm{k}}f_{z}(\bm{k})f_{z}(-\bm{k})
      +\mathcal{B}^{\mathrm{BA}}_{\bm{k}}f^{\dagger}_{xy}(\bm{k})d(\bm{k})
 \nonumber \\
 &\quad 
      +\mathcal{C}^{\mathrm{BA}}_{\bm{k}}f_{xy}(\bm{k})f_{xy}(-\bm{k})
      +\mathcal{D}^{\mathrm{BA}}_{\bm{k}}f_{xy}(\bm{k})d(-\bm{k})
      +\mathrm{H.c.}
     \biggr],
 \label{eq:QZ-BA}
\end{align}
where $\bar{\mathcal{F}}^{\mathrm{BA}}_{\mathrm{Q}}$ commutes with the
Hamiltonian~(\ref{eq:Hamiltonian-BA}), and the 
factors are 
$\mathcal{A}^{\mathrm{BA}}_{\bm{k}}=q/4E^{\mathrm{BA}_z}_{\bm{k}}$, 
$\mathcal{B}^{\mathrm{BA}}_{\bm{k}}=-\frac{\sin{2\theta}}{4}[\sqrt{\alpha_{\bm{k}}\beta_{\bm{k}}}+\frac{1}{\sqrt{\alpha_{\bm{k}}\beta_{\bm{k}}}}]$,
$\mathcal{C}^{\mathrm{BA}}_{\bm{k}}=-\frac{\tilde{q}}{4}[\alpha_{\bm{k}}-\frac{1}{\alpha_{\bm{k}}}]$,
$\mathcal{D}^{\mathrm{BA}}_{\bm{k}}=\frac{\sin{2\theta}}{4}[\sqrt{\alpha_{\bm{k}}\beta_{\bm{k}}}-\frac{1}{\sqrt{\alpha_{\bm{k}}\beta_{\bm{k}}}}]$,
$\alpha_{\bm{k}}=E^{\mathrm{BA}_{xy}}_{\bm{k}}/(\epsilon_{\bm{k}}+2n|c_1|)$, 
and
$\beta_{\bm{k}}=E^{\mathrm{BA_d}}_{\bm{k}}/\epsilon_{\bm{k}}$.
Equation~(\ref{eq:QZ-BA}) leads to the EAR
$R^{\mathrm{BA}}(\omega)=(\delta{q})^2\mathcal{R}_{\mathrm{c}}[r_{\mathrm{BA}_z}(\omega)+\theta_{\mathrm{H}}(|\tilde{\omega}|-2\tilde{\Delta}_{q}^{\mathrm{BA}})r_{\mathrm{BA}}(\omega)+\theta_{\mathrm{H}}(|\tilde{\omega}|-\tilde{\Delta}_{q}^{\mathrm{BA}})r_{\mathrm{BA}_{\mathrm{d}}}(\omega)]$ with 
\begin{align}
 r_{\mathrm{BA}_z}(\omega)
 &=\tilde{q}^2
   \sqrt{\frac{\sqrt{\tilde{q}^2+\tilde{\omega}^2}-\tilde{q}}{\tilde{q}^2+\tilde{\omega}^2}},
 \\
 r_{\mathrm{BA}_{xy}}(\omega)
 &=\tilde{q}^6
   \sqrt{\frac{\sqrt{\tilde{q}^4+\tilde{\omega}^2}+\tilde{q}^2-2}{\tilde{q}^4+\tilde{\omega}^2}},
 \\
 r_{\mathrm{BA}_\mathrm{d}}(\omega)
 &=\left(\frac{2|c_1|}{c_0}\right)^{3/2}
   \left(\tilde{\Delta}_{q}^{\mathrm{BA}}\right)^2
   |\tilde{\omega}|
 \nonumber \\
 &\quad
   \times
   \left(|\tilde{\omega}|-\tilde{\Delta}_{q}^{\mathrm{BA}}\right)^2
   \left[\gamma(\omega)-\frac{1}{\gamma(\omega)}\right]^2,
  \label{eq:EAR-BA}
\end{align}
where
$\gamma(\omega)=\sqrt{\frac{(c_0/|c_1|)^2\tilde{\Delta}_{q}^{\mathrm{BA}}}{(|\tilde{\omega}|-\tilde{\Delta}_{q}^{\mathrm{BA}})[(|\tilde{\omega}|-\tilde{\Delta}_{q}^{\mathrm{BA}})^2+c_0/|\tilde{c}_1|]}}$.
$2n|c_1|\tilde{\Delta}_{q}^{\mathrm{BA}}=\sqrt{(2nc_1)^2-q^2}$
denotes the energy gap of the spin mode,
$E^{\mathrm{BA}_{xy}}_{\bm{k}}$ (see Supplemental Material~\cite{sm}). 

The EAR spectra for the various $\tilde{q}$'s in the BA phase are shown in
Fig.~\ref{fig:EAR}(c). 
The gapless spectral weight $r_{\mathrm{BA}_z}(\omega)$ describes a
two-particle excitation of the gapless-spin mode
$E^{\mathrm{BA}_z}_{\bm{k}}$, and at the limit $\tilde{q}\rightarrow 1$,
it is identical to $R_{\mathrm{TP}}(\omega)$ at 
$\tilde{q}\rightarrow 1$.
The weight $r_{\mathrm{BA}_{xy}}(\omega)$, which is the two-particle
excitation of the spin mode $E^{\mathrm{BA}_{xy}}_{\bm{k}}$, gives
the gapful EAR spectrum with the gap, $2\tilde{\Delta}_{q}^{\mathrm{BA}}$, 
and vanishes at the limit $\tilde{q}\rightarrow 0$.
The other spectrum weight $r_{\mathrm{BA}_{\mathrm{d}}}(\omega)$ from
the pair excitation of the quasiparticles of the gapless-phonon mode
$E^{\mathrm{BA}_{\mathrm{d}}}_{\bm{k}}$ and of the gapful-spin
mode $E^{\mathrm{BA}_z}_{\bm{k}}$ provides a gapful spectrum
with the gap $\tilde{\Delta}_{q}^{\mathrm{BA}}$.
The two-particle excitation of the spin and density mode is peculiar to
the BA phase, as seen in the form of the QZ
perturbation~(\ref{eq:QZ-BA}).
In addition, the spectrum weight
$r_{\mathrm{BA}_{\mathrm{d}}}(\omega)$ is of the order of
$\sqrt{c_0/|c_1|}$, while the others are independent of $c_0$. 
Thus, the EAR for $c_0\gg |c_1|$ is dominated by
$r_{\mathrm{BA}_{\mathrm{d}}}(\omega)$ in the frequency regime where 
$r_{\mathrm{BA}_{\mathrm{d}}}(\omega)$ is finite.

\textit{Trap effect.---}
Based on the above results in the homogeneous case, we discuss
the EAR spectrum for trapped systems by local density approximation.
Then, the spectrum is calculated by taking the average of the local
spectrum with the weight of the local density $n(\bm{r})$.
The local EAR is obtained by replacing the mean density $n$
by the local one $n(\bm{r})$.
Since the density $n$ accompanies the interaction constant $c_0$ and
$c_1$ in all of the results, it turns out that the inhomogeneity
modifies the strength of the interactions: 
The effective interaction around the trap center is stronger at the
center, and gets weaker when going away from the center.
Thus, if the trap center is in the polar phase and in the FM, the whole
system exhibits a polar and FM state, respectively, as expected from
Fig.~\ref{fig:phase-diagram}. 
In addition, since the gap of the EAR spectrum is independent of the
density and the interaction, the gapful feature of the EAR in the polar
phases should remain even in the trapped case. 
On the other hand, when the trap center is in the BA phase, the outward
regime would be a polar state.
However, since the EAR in the BA phase is characterized by the gapless
spectrum, the qualitative feature is expected to be protected. 
Therefore, the EAR spectrum is concluded to characterize the phases of
spin-$1$ BECs regardless of the presence of trap potentials.

\textit{Effect of noncondensed fraction.---}
In spinor BECs, it may not be so trivial that the effect of 
noncondensed atoms is negligible.~\cite{Phuc.Kawaguchi.Ueda/PRA84.2011} 
Such a noncondensed fraction can be regarded as some kind of
fluctuation, and the Bogoliubov spectra is thus expected to capture the
physics.
Since the main features of the obtained spectra come from the
spectral character of the fluctuations, they should be visible
even in the presence of the noncondensed fraction.
Therefore, if the system is cooled down enough
such that the temperature is less than
the mean-field energy,
the spectra demonstrated here would be measured
due to the bosonic stimulation effect.

\textit{Conclusion.---}
We have formulated the response of the spinor BECs to modulation of the
magnetic field by using linear response theory, which gives access to
the correlation of the QZ term, and, by considering the spin-$1$ BECs,
the spectrum has been demonstrated to have individual features in
each magnetic phase.
In addition, the results have been found robust against the trap effect.

Finally, we comment on potential applications of this spectroscopy. 
From the high versatility of the formula~(\ref{eq:EAR}), it can be
widely applied, for example, to BECs with any spin, high-spin Fermi
atoms, and optical lattice systems.
Furthermore, it would also be used for the following fundamental
problems: an experimental test to verify Bogoliubov theory for spinor
BECs, and dimensionality discussion on single- and multispatial spin
modes for an anisotropic trap by using the fact that the different spectral
shape strongly depends on the system dimensions.

\acknowledgements
We thank Yuki Kawaguchi for fruitful comments and Thierry Giamarchi for
critical reading of the manuscript. 
This work was supported by the Swiss National Science Foundation under
MaNEP and division II.

\bibliographystyle{apsrev4-1}

%

\end{document}